\documentclass{llncs}
\usepackage{makeidx}  
\usepackage{graphicx}
\usepackage{xcolor}

\begin{document}
\frontmatter          
\pagestyle{headings}  
\addtocmark{Hamiltonian Mechanics} 
\mainmatter              
\title{A Taxonomy of Hyperlink Hiding Techniques}
\titlerunning{Hamiltonian Mechanics}  
%
\author{Guang-Gang Geng\inst{1} \and  Xiu-Tao Yang\inst{2} \and Wei Wang\inst{1} \and Chi-Jie Meng\inst{1}}
\authorrunning{Guang-Gang Geng et al.} 
%
\tocauthor{Guang-Gang Geng, Xiu-Tao Yang, Wei Wang and Chi-Jie Meng}
\institute{China Internet Network Information Center,
Computer Network Information Center, Chinese Academy of Sciences, Beijing, China, 100180\\
\email{\{gengguanggang,wangwei,mengchijie\}@cnnic.cn}
\and
Beijing Institute of Electronic System Engineering, Beijing, China, 100854\\
\email{xiutaoyang\_temp@163.com}
}

\maketitle              

\begin{abstract}
Hidden links are designed solely for search engines rather than visitors. To get high search engine rankings, link hiding techniques are usually used for the profitability of underground economies, such as illicit game servers, false medical services, illegal gambling, and less attractive high-profit industry.
This paper investigates hyperlink hiding techniques on the
Web, and gives a detailed taxonomy. We believe the taxonomy
can help develop appropriate countermeasures.

Statistical experimental results on real Web data indicate that link hiding techniques are very
prevalent. We also tried to explore the attitude of Google towards link hiding spam by analyzing the PageRank values of relative links. The results show that more should be done to punish the hidden link spam.
\end{abstract}
\section{Introduction}
Most Web surfers depend on search engines to locate information on the Web. Link analysis algorithms \cite{ng2001stable}, such as PageRank \cite{page1999pagerank} and HITS \cite{kleinberg1999authoritative}, are usually used for Search engines ranking. Link analysis algorithms assume that every link represents a vote of support, in the
sense that if there is a link from page x to page y and these two pages are authored by different people, then the author of page x is
recommending page y.
In particular,
PageRank is the basis of Google's search technology \cite{brin1998anatomy}.

Web spammers try to mislead search engines to make a high rank in search results \cite{gyongyi2005web}.
In this context, hyperlink hiding techniques are often used to deceive search engines. Spammers hope that many small endorsements from these pages with hidden links result in a sizable PageRank for the target page. Several questions naturally arise: what link hiding techniques are the spammers using; and, how prevalent are hidden spam links on the Web?
This paper attempts to answer those questions.

The rest of sections are organized as follows. Section 2 presents a literature review. Section 3 gives a comprehensive taxonomy of current hidden link spam techniques. Section 4 describes the experimental analysis on 5,583,451 Chinese Web sites. At last, section 5 draws the conclusion.

\section{Related Work}
Hidden links are designed to increase link popularity, which are invisible for visitors
\cite{wiki:hiddenlink}. Google considers hyperlinks hidden by small characters as deception \cite{google}. Gyongyi etc al. point out that hidden links are often used in honey pot to boost the ranking of the spam pages \cite{gyongyi2004combating}. They further present a comprehensive taxonomy of current spamming techniques and survey content hiding techniques, where spam links hidden by avoiding anchor texts or tiny anchor images are mentioned \cite{gyongyi2005web}. Link-hiding related features are not paid more attention to in statistical Web spam detection studies \cite{erdelyi2011web}\cite{liu2012identifying}\cite{spirin2012survey}.


To the best of our knowledge, there is no previously published literature that
directly studied how prevalent, successful, or varied hidden link spam techniques are on the Web.
This paper attempts to study
hidden link spam in detail. It is hoped that the findings
can help in developing appropriate countermeasures.

\section{Hyperlink Hiding Techniques}\label{tech}
There are many different ways to hide links from visitors while leaving it perfectly viewable to search engines.
In this section, we will examine current hyperlink hiding techniques used by spammers and attempt to
categorize them based on their features. Just as the work on JavaScript redirection spam \cite{chellapilla2007taxonomy}, we present short examples to show the hiding techniques really used by spammers. Simple techniques are presented first and are
followed by more advanced ones.

\subsection{A: Making Anchor Text Font Color the Same as Background Color}
The simplest and oldest method that spammers use to create hidden links is to make the font of anchor text the same color as the background.
Here is one example.
\renewcommand\arraystretch{1}
\vspace*{-10pt}

\begin{table}[!htbp]
\centering
\colorbox[rgb]{0.9,0.9,0.9}{%
\begin{tabular}[!htp]{l}

$<$span style=$``$background$:$white$;">$ \\
\ \ \ \ $<$a href=$``$target.html$"$ style=$``$color:white$"$$>$ {invisible anchor text} $<$$/$a$>$ \\
$<$$/$span$>$ \\

\end{tabular}}\label{tb1}
\end{table}
\vspace*{-10pt}

In this example, the color scheme is defined in the HTML document. Color schemes can also be defined in an attached cascading style sheet file (CSS). Sometimes, spammers also consider background images. They set the image color to be the same as the font color, which is relatively harder to detect.

\subsection{B: Making Anchor Text Font Color Almost Match Background Color or Background Image}
Instead of setting the font color to entirely match the background color, some spammers and web masters set their font colors to almost match the background color. The idea behind this method is that they believe that they are thwarting the search engines' software detection systems by slightly changing the color of the text.

\renewcommand\arraystretch{1}
\vspace*{-10pt}
\begin{table}[!htbp]
\centering
\colorbox[rgb]{0.9,0.9,0.9}{%
\begin{tabular}[!htp]{l}
$<$div style=$``$background-color$:$white$;">$ \\
\ \ \ \ $<$a href=$``$target.html$"$ style=$``$color:$\#$feffee$"$$>$ {text color similar to white} $<$$/$a$>$ \\
$<$$/$div$>$ \\

\end{tabular}}\label{tb1}
\end{table}
\vspace*{-10pt}

\subsection{C: Setting Tiny Anchor Text or Placing the hyperlinks in a Tiny Block}
Making tiny anchor text is another hyperlink hiding method. This way, the hyperlink can be set small enough, such as 1 pixel high, even 0 pixel.
Here's a simple example of that.

\renewcommand\arraystretch{1}
\vspace*{-10pt}
\begin{table}[!htbp]
\centering
\colorbox[rgb]{0.9,0.9,0.9}{%
\begin{tabular}[!htp]{l}
\ \ \ \ $<$a href=$``$target.html$"$ style=$``$font-size:0px$"$$>$ {tiny text} $<$$/$a$>$
\end{tabular}}\label{tb2}
\end{table}
\vspace*{-10pt}

In HTML, the div element is often used for generic organizational or stylistic applications. Spammers can also use div to set the link size. The following is another example.

\renewcommand\arraystretch{1}
\vspace*{-20pt}
\begin{table}[!htbp]
\centering
\colorbox[rgb]{0.9,0.9,0.9}{%
\begin{tabular}[htp]{l}
$<$div style=$``$font-size:0px$;$$"$$>$ $<$a href=$``$target.html$"$ $>$invisible text$<$$/$a$>$ $<$$/$div$>$ \\
\end{tabular}}\label{tb3}
\end{table}
\vspace*{-20pt}

Perhaps the most common use of div element is to carry class or id attributes in conjunction with CSS to apply layout, typographic, color, and other presentation attributes to parts of the content. In the previous example, the \emph{font-size:0px} can also be defined in a CSS file. Besides, div block size can be set via width and height attributes. For example, $<$div style=$``$width:1px;height:1px$;$$"$$>$, where the div size is 1 pixel.

Another example of hiding a hyperlink via tiny scrolling block is presented below.
\renewcommand\arraystretch{1}
\vspace*{-20pt}
\begin{table}[!htpb]
\centering
\colorbox[rgb]{0.9,0.9,0.9}{%
\begin{tabular}[htp]{l}
$<$marquee scrollAmount=1 width=1 height=1$>$\\
\ \ \ \ $<$a href=$``$target.html$"$$>$ \emph{text in a tiny scrolling block} $<$/a$>$\\
$<$$/$marquee$>$
\end{tabular}}\label{tb13}
\end{table}
\vspace*{-20pt}

In this example, $target.html$ is put in a scrolling block with area $1\times{1}$ pixel, which is invisible to Web users.

\subsection{D: Disguising Anchor Text as Plain Text}
Sometimes, spammers insert hyperlinks into a paragraph, where the anchor text looks like plain text.
Here's a paragraph of text on a site:

\vspace*{-10pt}
\begin{figure}[!htp]\centering
\setlength{\fboxrule}{0.5pt}
\setlength{\fboxsep}{0.05cm}
\fbox{\includegraphics[angle=0, width=0.6\textwidth]{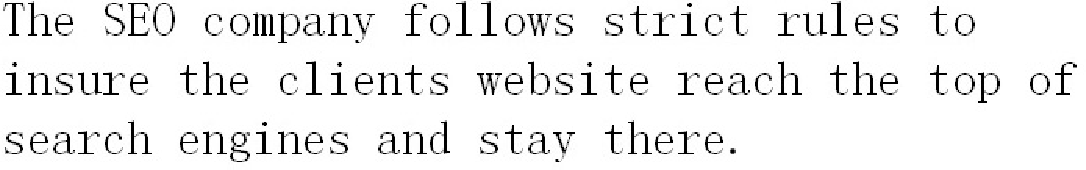}}
\end{figure}
\vspace*{-10pt}

A user wouldn't see any hyperlinks, even if they moused over every word in the paragraph. But if you happened to click on just the right word, you'd get whisked away to a SEO site. Actually, there is a hidden link under the anchor text ``SEO company". If you view the source of the page, here's what you'll see:

\vspace*{-10pt}
\begin{table}[!htbp]
\centering
\colorbox[rgb]{0.9,0.9,0.9}{%
\begin{tabular}[htp]{l}
The $<$a href=$``$http$:$$/$$/$www.seomarketleaders.com$"$ onMouseOver=\\$``$window.status=`';return true;$"$ style=$``$cursor$:$text;color:black; \\text-decoration:none;$"$$>$ SEO company$<$$/$a$>$ follows strict \\rules to insure the clients website reach the top of search engines and stay there.
\end{tabular}}\label{tb101}
\end{table}
\vspace*{-10pt}


\subsection{E: Placing Hyperlinks in High-Speed Scrolling Blocks}
The $<$marquee$>$ tag is a non-standard HTML element which causes text to scroll up, down, left or right automatically \cite{wiki:Marquee}. Although the W3C advises against its use in HTML documents, it's still widely used. SCROLLAMOUNT attribute sets the speed of the scrolling. A bigger value for SCROLLAMOUNT makes the marquee scroll faster. If the SCROLLAMOUNT value is big enough, the scrolling block will be invisible to the naked eye. Here is a simple example.

\renewcommand\arraystretch{1}
\vspace*{-10pt}
\begin{table}[!htp]
\centering
\colorbox[rgb]{0.9,0.9,0.9}{%
\begin{tabular}[htp]{l}
$<$marquee height=1 width=8 scrollamount=3000$>$ \\
\ \ \ \ $<$a href=$``$target.html$"$$>$ \emph{text in a high-speed scrolling block} $<$$/$a$>$\\
$<$$/$marquee$>$
\end{tabular}}\label{tb3}
\end{table}
\vspace*{-10pt}

The default $scrollamount$ value is 6. The value in the example is 3000, which is too fast to see.

Similar effects can also be achieved through the use of JavaScript or HTML $<$blink$>$ element \cite{wiki:Marquee} \cite{wiki:Blink}.

\subsection{F: Putting Links outside the Screen}

Using cascading style sheets, you have the option to absolutely or relatively position any division. Using absolute position, you can simply position the text you wish to hide any number of pixels off the screen to the left of the window. Here are some example codes:

\renewcommand\arraystretch{1}
\vspace*{-20pt}
\begin{table}[!htp]
\centering
\colorbox[rgb]{0.9,0.9,0.9}{%
\begin{tabular}[htp]{l}
$.$hiddenclass $\{$ position $:$ absolute;left $:$ -977px; $\}$
\end{tabular}}\label{tb3}
\end{table}
\vspace*{-20pt}

If you put that in your style sheet and then assign the class ``hiddenclass" to your div, then the div will display 977 pixels to the left of the visible screen - i.e., it will not appear on the screen. Here is a example:
\vspace*{-20pt}
\begin{table}[!htp]
\centering
\colorbox[rgb]{0.9,0.9,0.9}{%
\begin{tabular}[htp]{l}
$<$div id=$``$hiddenclass$"$$>$ $<$a href=$``$target.html$"$$>$ \emph{invisible anchor text} $<$$/$a$>$ $<$$/$div$>$\\
\end{tabular}}\label{tb3}
\end{table}
\vspace*{-20pt}

The absolute position can also be set in the div directly as follows:
\renewcommand\arraystretch{1.5}
\vspace*{-10pt}
\begin{table}[!htp]
\centering
\colorbox[rgb]{0.9,0.9,0.9}{%
\begin{tabular}[htp]{l}
$<$div style=$``$left$:$ -977px; position: absolute; top: -977px$"$$>$ \\
\ \ \ \ $<$a href=$``$target.html$"$$>$ \emph{invisible anchor text} $<$$/$a$>$\\
$<$$/$div$>$\\
\end{tabular}}\label{tb3}
\end{table}
\vspace*{-10pt}

In the example above, $left:-977$ may be written in more complex formats, such as $left:expression(23-1000)$.

In addition to the methods described above, users can use CSS text-indent property or margin-left property
to put hyperlinks outside the screen. A example is presented below.

\renewcommand\arraystretch{1}
\vspace*{-10pt}
\begin{table}[!htp]
\centering
\colorbox[rgb]{0.9,0.9,0.9}{%
\begin{tabular}[htp]{l}
$<$div style=$``$text-indent:-999px;$"$$>$$<$a href=$``$target.html$"$$>$\emph{hidden text}$<$/a$>$ $<$$/$div$>$
\end{tabular}}\label{tb3}
\end{table}
\vspace*{-20pt}

\subsection{G: Using Visibility:Hidden or Display:None Style Commands}\label{vis}
An alternative to the method above is to simply use the built in features of style sheets to hide hyperlinks:
\renewcommand\arraystretch{1}
\begin{table}[!htp]
\centering
\colorbox[rgb]{0.9,0.9,0.9}{%
\begin{tabular}[htp]{l}
$.$hiddenclass $\{$ visibility $:$ hidden;$\}$
\end{tabular}}\label{tb3}
\end{table}

Again, if you put that into a style sheet and then assign the class ``hiddenclass" to your div, the hyperlinks in the div block will not appear in the browser window.

\subsection{H: Hiding Hyperlinks via JavaScript}
JavaScript is an open source programming language commonly implemented as part of a web browser in order to create enhanced user interfaces and dynamic websites \cite{flanagan2006javascript}. Google claims that search engines have difficulty accessing JavaScript \cite{google}. In 2011, labnol.org reported that Google indexes JavaScript based Facebook comments, but there is no clear report that Google parsers JavaScript codes on the whole Web. This fact encourages spammers to hide hyperlinks by the aid of JavaScript. Here is a simple example:

\renewcommand\arraystretch{1}
\vspace*{-10pt}
\begin{table}[!htbp]
\centering
\colorbox[rgb]{0.9,0.9,0.9}{%
\begin{tabular}[htp]{l}
$<$script language=$`$$`$JavaScript$"$ type=$``$text/javascript$"$$>$\\
\ \ \ \ document.write( $``$$<$div style=$`$visibility:hidden'$>$$"$ );\\
$<$/script$>$ \\
$<$a href=$``$target.html$"$$>$\emph{keywords}$<$$/$a$>$\\
$<$script language=$`$$`$JavaScript$"$ type=$``$text/javascript$"$$>$ \\
\ \ \ \ document.write( $`$$`$$<$/div$>$$"$ ); \\
$<$/script$>$
\end{tabular}}\label{tb3}
\end{table}
\vspace*{-10pt}

The example is easy to understand, which is a packaging of the method described in section \ref{vis}. In the above codes, $<$div$>$ and $<$$/$div$>$ tags are embed in JavaScript codes separately, which may not be indexed by search engines. However, the hyperlink \emph{target.html} is displayed in html codes, which is more likely to indexed by search engines.
In a similar manner, almost all the link hiding techniques described in this section can be further disguised with JavaScript. Next, let's look into a more complex example.

\renewcommand\arraystretch{1}
\begin{table}[!htbp]
\centering
\colorbox[rgb]{0.9,0.9,0.9}{%
\begin{tabular}[htp]{l}
$<$div id=$``$$ql1000$$"$$>$ \\
\ \ \ \ $<$a href=$``$target.html" title=$``$keyword$"$$>$\\
\ \ \ \ \ \ \ \ \emph{target keyword}\\
\ \ \ \ $<$$/$a$>$\\
$<$$/$div$>$\\
$<$script language=$``$JavaScript$"$$>$\\
var $\_xa$= $[$\\
\ \ \ \ \footnotesize{$``$$\backslash$x64$\backslash$69$\backslash$x73$\backslash$x70$\backslash$x6C$\backslash$x61$\backslash$x79$"$$,$
$``$$\backslash$x6E$\backslash$x6F$\backslash$x6E$\backslash$x65$"$$,$}\\
\ \ \ \
\footnotesize{$``$$\backslash$x71$\backslash$x6c$\backslash$x31$\backslash$x30$\backslash$x30$\backslash$x30$"$$,$
$``$$\backslash$x73$\backslash$x74$\backslash$x79$\backslash$x6C$\backslash$x65$"$$,$}\\
\ \ \ \ \footnotesize{$``$$\backslash$x67$\backslash$x65$\backslash$x74$\backslash$x45$\backslash$x6C$\backslash$x65$\backslash$x6D$\backslash$x65$\backslash$x6E$\backslash$x74$\backslash$x42$\backslash$x79$\backslash$x49$\backslash$x64$"$$]$;}\\
\ \ \ \ document$[$$\_xa$$[$4$]$$]$($\_xa$$[$2$]$)$[$$\_xa$$[$3$]$$]$$[$$\_xa$$[$0$]$$]$=$\_xa$$[$1$]$;\\
$<$/script$>$
\end{tabular}}\label{tb70}
\end{table}

The above JavaScript codes are designed in rather vague terms. The elements of array $\_xa$ are written with ASCII characters.
The last line of the above JavaScript codes is
document$[$$`$getElementById'$]$$($$`$ql1000'$)$$[$$`$style'$]$$[$$`$display'$]$=$`$none',
which makes all the content, including hyperlinks, in the div named $ql1000$ invisible.
In order to avoid presenting the whole style assignment directly, 
script can build up the style assignment via string concatenation.
One very straight forward example is presented below.

\renewcommand\arraystretch{1}
\vspace*{-10pt}
\begin{table}[!htp]
\centering
\colorbox[rgb]{0.9,0.9,0.9}{%
\begin{tabular}[htp]{l}
$<$script
type=$`$$`$text$/$javascript$"$$>$\\
\ \ \ \ document.getElementById($`$$`$q$"$ + $`$$`$l$"$ +
$`$$`$1000$"$).style.display=$`$$`$n$"$ + $`$$`$o$"$ +
$`$$`$ne$"$;\\
$<$$/$script$>$
\end{tabular}}\label{tb8}
\end{table}
\vspace*{-10pt}

What is worse, JavaScript as a programming language, has many functions and operators, which throw off a human
readers. The following codes show the flexibility of JavaScript.
\renewcommand\arraystretch{1}
\vspace*{-20pt}
\begin{table}[!htp]
\centering
\colorbox[rgb]{0.9,0.9,0.9}{%
\begin{tabular}[htp]{l}
$<$script language=$`$$`$javascript$"$$>$function HexTostring(s)\{\\
\ \ \ \ var r=`';\\
\ \ \ \ for(var i=0;i$<$s.length;i+=2)\{\\
\ \ \ \ \ \ \ \ var sxx=parseInt(s.substring(i,i+2),16);\\
\ \ \ \ r+=String.fromCharCode(sxx);\}\\
\ \ \ \ return r;\}\\
\ \ \ \ eval(HexTostring($`$$`$646f63756d656e742e676574456c656d65\\
\ \ \ \ 6e74427949642822716c3130303022292e7374796c652e6469\\
\ \ \ \ 73706c6179203d20226e6f6e6522$"$));\\
$<$/script$>$
\end{tabular}}\label{tb17}
\end{table}
\vspace*{-20pt}

These codes are essentially equivalent to the previous example, yet look completely different.

\subsection{I: Hiding Hyperlinks via Cloaking or Redirection Techniques}
Cloaking is a Web spam technique in which the page presented to the
search engine spider is different from that presented to the user's
browser \cite{wu2005cloaking}. Some spammers hide target
hyperlinks using cloaking technique. Similarly, spammers also use
redirection techniques to hide targeting hyperlinks. Among the
redirection spam techniques, JavaScript based redirection is the
most notorious and difficult to catch
\cite{chellapilla2007taxonomy}. Wu et al. \cite{wu2005cloaking} and
Chellapilla et al. \cite{chellapilla2007taxonomy} have conducted
comprehensive studies of cloaking and redirection techniques
respectively, so the techniques will not be repeated here. However,
it's important to point out that we do not consider the redirected
target URL, but the hyperlinks in the redirection page as hidden
links. For example, $A$ redirects to $B$, and $C$ is a hyperlink in
page $A$. In this paper, $C$ is a hidden link, but $B$ is not seen
as a hidden link.

\subsection{J: Hiding Hyperlinks in Pull-Down Menu}
Pull-down menu is also called a drop-down menu, which is a menu of commands or
options that appears when you select an item with a mouse. A drop
down menu can make it easier to display a large list of choices -
since only one choice is displayed initially, the remaining choices
can be displayed when the user activates the dropbox. Some spammers
insert the target hyperlinks into a long pull-down list, which are
hard to find.

\subsection{K: Inserting Links into Long Title or Meta Tags}
Generally, web browsers show the preceding part of a long title. Thus, some spammers insert urls into long title.
Similarly, meta tags provide structured meta data
about a Web page and they are used for search engines. Although they have been the targets of
spammers for a long time and search engines consider
these data less and less, there are pages still using
them.

\subsection{L: Hiding Div ``Below" the Visible Layer}
Another sneaky way to hide a hyperlink from Web users while keeping
it available to the search engines is to put the hyperlinks in a
layer that is ``behind" the visible layer. The CSS z-index property
specifies the stack order of an element, which is supported in all
major browsers. An element with a greater stack order is always in
front of an element with a lower stack order. One example hiding
hyperlinks via z-index is presented below.
\renewcommand\arraystretch{1}
\begin{table}[htbp]
\centering
\colorbox[rgb]{0.9,0.9,0.9}{%
\begin{tabular}[!htp]{l}
$<$div id=$`$$`$front$"$ style=$`$$`$position:absolute; z-index:1$"$$>$\\
\ \ \ \ $<$img src=$`$$`$$image.gif$$"$ $>$\\
$<$$/$div$>$ \\
$<$div id=$`$$`$back$"$ style=$`$$`$position:absolute; z-index:-1$"$$>$\\
\ \ \ \ $<$a href=$`$$`$$target.html$$"$ target=$`$$`$$\_$blank$"$$>$ \emph{target keyword}$<$/a$>$\\
$<$$/$div$>$
\end{tabular}}\label{tb9}
\end{table}

The codes show that the second div has a negative stack order, which
determines the $target.html$ is behind the $image.gif$. Besides z-index, ``overflow:hidden" can also hide the hyperlinks below the visible layer. Here is a simple example.

\renewcommand\arraystretch{1.2}
\begin{table}[htbp]
\centering
\colorbox[rgb]{0.9,0.9,0.9}{%
\begin{tabular}[!htp]{l}
$<$style type=$`$$`$text/css$"$$>$\\
\#spam\{width:99px;height:20px;overflow:hidden;position:absolute;\}\\
\#spam a\{display:block;line-height:20px;text-decoration:none;\}\\
$<$/style$>$\\
$<$div id=$`$$`$spam$"$$>$\\
\ \ \ \ $<$a href=$`$$`$/$"$$>$\&\#160;$<$/a$>$\\
\ \ \ \ $<$a href=$`$$`$$target.html$$"$ title=$`$$`$keywords$"$$>$\\
\ \ \ \ \ \ \ \ \emph{target keyword}\\
\ \ \ \ $<$/a$>$\\
$<$$/$div$>$
\end{tabular}}\label{tb11}
\end{table}
In the example above, $target.html$ is covered by a non-breaking space.

\section{Prevalence of Link Hiding Techniques}
Link-hiding can be considered an adversarial problem. As commercial search engines develop algorithms to detect and discard certain types of hidden links, new techniques for hiding links will be developed. In last section, we examined current hyperlink hiding techniques used by spammers and categorized them based on their features. In this section, we study the prevalence of hidden spam links, and how prevalence of the variety of techniques
described in Section \ref{tech}. 

We carried out the analysis on 5,765,357 Chinese homepages (\emph{http://www.} + \emph{domain name}) in Sep. 2012, including .com, .net and .cn domain names. To detect the Web pages with hiding links, we first train a cost sensitive naive bayes classifier on 103 pages with hidden links and 271 normal pages. The cost sensitive model ensures a high recall of pages with hidden links. Then, we filtered the 5,765,357 pages with the trained model. The detection results contain quite a few false alarms, but it's enough for us to analyze the prevalence of hidden links. By random sampling from the suspicious set and carrying out manual verification, we approximately determined the number of pages with hidden links. Table \ref{tbr1} tabulates the statistics in detail.

\begin{table}[htbp]
 \caption{Percentage occurrence of hidden link spam
among Chinese Web pages }
\centering
\tabcolsep 0.02in
\begin{tabular}[htp]{c|c|c}
\hline \hline
URL Type & Count / Total & Percentage\\
\hline
.com/.net/.cn & $81775/5765357=1/70.5$ & 1.42\%\\
\hline
\end{tabular}\label{tbr1}
\end{table}
It is noticed that a number of Chinese pages use hyperlink hiding techniques. 
To analyze the prevalence of the variety of techniques described in Section \ref{tech}, we randomly sampled 4727 pages with hidden links from .com/.net/.cn set. Each sampled hidden link spam page was manually analyzed. All the 4727 samples were labeled with the types of techniques they used. Besides, all the hidden links are extracted for further analysis. In total, 16767 unique target hyperlinks are hidden in the 4727 pages.

Table \ref{tbr2} describes the prevalence of hidden link techniques in detail. The table shows that the 4727 pages contain 16767 unique hidden links. F, G and H are the most popular link hiding techniques, which account for 75.3\% of that total. These three techniques can be easily used to hide multiple hyperlinks. It can be observed that some of the 4727 web pages contain more than one link hiding technique.
\begin{table}[htbp]
 \caption{Prevalence of different link hiding techniques }
\centering
\begin{tabular}[htp]{c|c}
\hline \hline
{techniques} & {number(percentage)$=>$number of hidden links}\\
\hline
A & 102 (2.2\%) $=>$ 661 \\
B & 51 (1.1\%) $=>$ 493 \\
C & 137 (2.9\%) $=>$ 561 \\
D & 322 (6.8\%) $=>$ 357 \\
E & 136 (2.9\%) $=>$ 1987 \\
F & 1157 (24.5\%) $=>$ 68661\\
G & 511 (10.8\%) $=>$ 30192 \\
H & 1888 (39.9\%) $=>$ 51570 \\
I & 86 (1.8\%) $=>$ 2071 \\
J & 151 (3.2\%) $=>$ 527 \\
K & 255 (5.4\%) $=>$ 103 \\
L & 53 (1.1\%) $=>$ 779 \\
\hline
{All} & {4849 (4849/4727=102.6\%) $=>$ 157962(unique links: 16767)}\\
\hline
\end{tabular}\label{tbr2}
\end{table}

Are the 16767 target pages punished by the search engines? We do not know the detailed ranking strategy of commercial search engines, but we can explore this problem from a side by analyzing the PageRank values of the target hyperlinks. Google provides a public interface, toolbarqueries.google.com, for querying the PageRank values. Table \ref{tbr3} shows the average PageRank values of target hidden links and randomly selected 39756 urls from DNS resolution logs.

\begin{table}[htbp]
\caption{Comparison of average PageRank values}
\centering
\begin{tabular}[htp]{c|c|c}
\hline \hline
 & hidden links&randomly selected urls\\
\hline
Number& 16767 & 39756\\
\hline
Average PageRanks& 1.340 & 1.137 \\
\hline
\end{tabular}\label{tbr3}
\end{table}

Table \ref{tbr3} shows that the 16767 hidden links have an average PageRank value 1.34, which is higher than that of the randomly selected urls. To some extent, the result means that Google needs to establish a more effective punitive mechanism for the hidden links.

\begin{figure}[htbp]
\centerline{\includegraphics[scale=0.44]{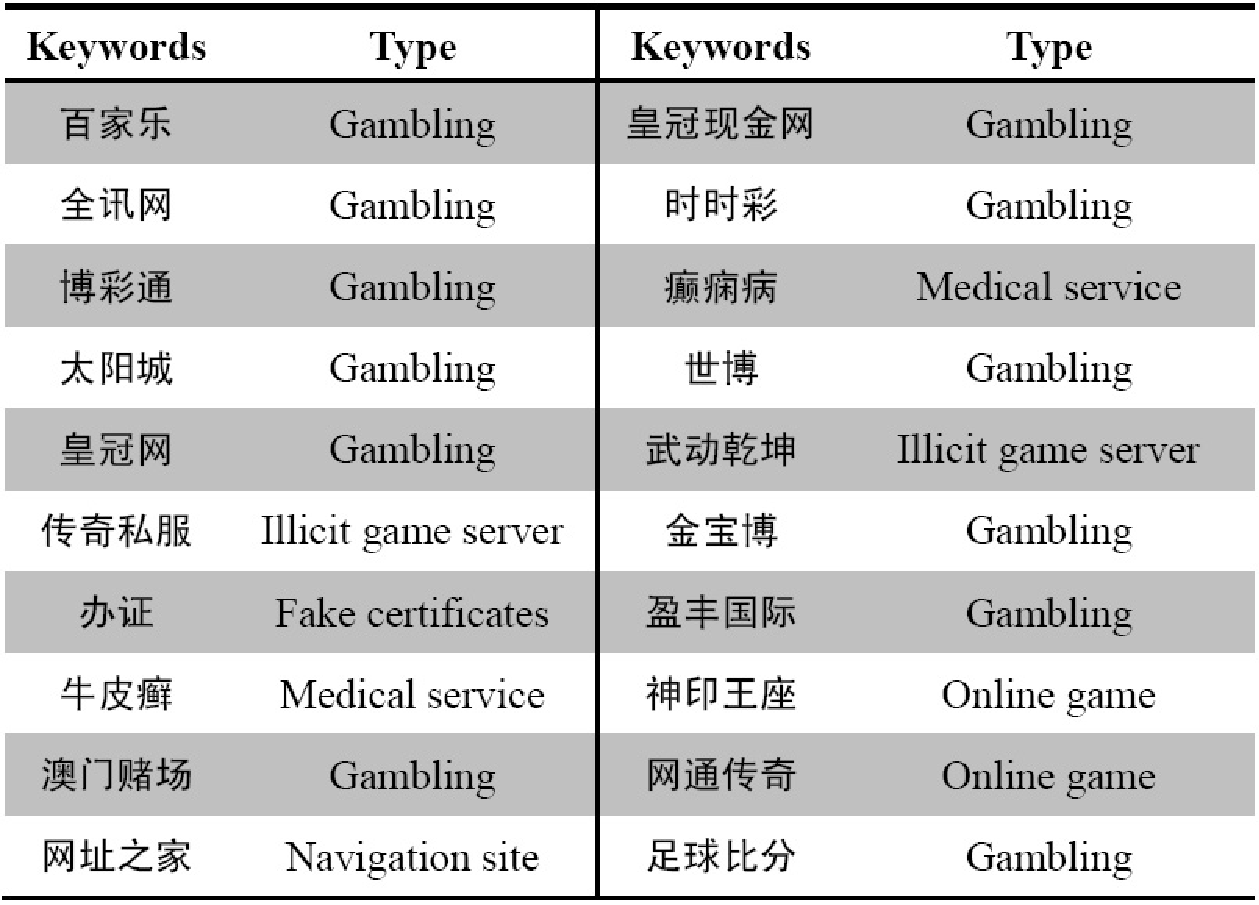}}
\caption{The high-frequency words in the anchor texts of the target hyperlinks.} \label{fig_key}
\end{figure}

We further analyzed the high-frequency words in the anchor texts of the 16767 target hyperlinks. The top 20 high-frequency keywords and the corresponding types are described in figure \ref{fig_key}.
The statistics show that gambling sites, personal game servers and medical services are the main types of the hidden links. Most of the sites belong to shady or illegal industries.

\section{Conclusion and Future Work}
In this paper we presented a variety of commonly used
link hiding techniques, and organized them into a
taxonomy. We analyzed the prevalence of common link hiding techniques on the
web.
Just as the previous work on Web spam \cite{gyongyi2005web} \cite{chellapilla2007taxonomy}, we argue that such a structured discussion
of the subject is important to raise the awareness of
the research community. Given that most of the sites using link hiding techniques are shady or illegal industries, more should be done to punish the hidden link spam.

In the future, we should pay more attention to two things. The first is studying link hidden spam on a bigger data set, which includes multilingual samples. The second is developing a proper countermeasure to address the problem as a whole, despite the variety of different link hiding techniques. One possible solution draws support from maturing optical character recognition techniques (OCR) \cite{mori1999optical}. The motivation is that as a computer vision technique, OCR can only read the visible content on the Web page like humans. The snapshot of a Web page can be easily taken via some softwares, such as wkhtmltopdf \cite{wkhtmltopdf} and snapshotter \cite{Snapshotter}. All the visual text on the snapshot image can be recognized via OCR techniques as $textVector$. If an anchor text does not exist in the $textVector$, the corresponding hyperlink is identified as hidden link. And, of course, the relative position of anchor text should also be taken into account.

\section*{Acknowledgment}

This paper is supported by
grants National Natural Science Foundation of
China (Nos. 61005029, 61103138 \& 61375039). This paper is the result of many beneficial discussions with one of our collaborators at a major search engine company, who wishes to remain anonymous.

\bibliographystyle{IEEEtran}

\bibliography{IEEEabrv,hid}

\begin{thebibliography}{10}
\providecommand{\url}[1]{#1}
\csname url@samestyle\endcsname
\providecommand{\newblock}{\relax}
\providecommand{\bibinfo}[2]{#2}
\providecommand{\BIBentrySTDinterwordspacing}{\spaceskip=0pt\relax}
\providecommand{\BIBentryALTinterwordstretchfactor}{4}
\providecommand{\BIBentryALTinterwordspacing}{\spaceskip=\fontdimen2\font plus
\BIBentryALTinterwordstretchfactor\fontdimen3\font minus
  \fontdimen4\font\relax}
\providecommand{\BIBforeignlanguage}[2]{{%
\expandafter\ifx\csname l@#1\endcsname\relax
\typeout{** WARNING: IEEEtran.bst: No hyphenation pattern has been}%
\typeout{** loaded for the language `#1'. Using the pattern for}%
\typeout{** the default language instead.}%
\else
\language=\csname l@#1\endcsname
\fi
#2}}
\providecommand{\BIBdecl}{\relax}
\BIBdecl

\bibitem{ng2001stable}
A.~Ng, A.~Zheng, and M.~Jordan, ``Stable algorithms for link analysis,'' in
  \emph{Proceedings of the 24th annual international ACM SIGIR conference on
  Research and development in information retrieval}.\hskip 1em plus 0.5em
  minus 0.4em\relax ACM, 2001, pp. 258--266.

\bibitem{page1999pagerank}
L.~Page, S.~Brin, R.~Motwani, and T.~Winograd, ``The pagerank citation ranking:
  bringing order to the web.'' 1999.

\bibitem{kleinberg1999authoritative}
J.~Kleinberg, ``Authoritative sources in a hyperlinked environment,''
  \emph{Journal of the ACM (JACM)}, vol.~46, no.~5, pp. 604--632, 1999.

\bibitem{brin1998anatomy}
S.~Brin and L.~Page, ``The anatomy of a large-scale hypertextual web search
  engine,'' \emph{Computer networks and ISDN systems}, vol.~30, no.~1, pp.
  107--117, 1998.

\bibitem{gyongyi2005web}
Z.~Gyongyi and H.~Garcia-Molina, ``Web spam taxonomy,'' in \emph{First
  international workshop on adversarial information retrieval on the web
  (AIRWeb 2005)}, 2005.

\bibitem{wiki:hiddenlink}
\BIBentryALTinterwordspacing
Wikipedia, ``Spamdexing --- {W}ikipedia{,} the free encyclopedia,'' 2013,
  [Online; accessed 17-January-2013]. [Online]. Available:
  \url{\url{http://en.wikipedia.org/wiki/Spamdexing}}
\BIBentrySTDinterwordspacing

\bibitem{google}
\BIBentryALTinterwordspacing
Google, ``Webmaster guidelines - webmaster tools help,'' 2013, [Online;
  accessed 17-January-2013]. [Online]. Available:
  \url{http://www.google.com/webmasters/guidelines.html}
\BIBentrySTDinterwordspacing

\bibitem{gyongyi2004combating}
Z.~Gy{\"o}ngyi, H.~Garcia-Molina, and J.~Pedersen, ``Combating web spam with
  trustrank,'' in \emph{Proceedings of the Thirtieth international conference
  on Very large data bases-Volume 30}.\hskip 1em plus 0.5em minus 0.4em\relax
  VLDB Endowment, 2004, pp. 576--587.

\bibitem{erdelyi2011web}
M.~Erd{\'e}lyi, A.~Garz{\'o}, and A.~A. Bencz{\'u}r, ``Web spam classification:
  a few features worth more,'' in \emph{Proceedings of the 2011 Joint
  WICOW/AIRWeb Workshop on Web Quality}.\hskip 1em plus 0.5em minus 0.4em\relax
  ACM, 2011, pp. 27--34.

\bibitem{liu2012identifying}
Y.~Liu, F.~Chen, W.~Kong, H.~Yu, M.~Zhang, S.~Ma, and L.~Ru, ``Identifying web
  spam with the wisdom of the crowds,'' \emph{ACM Transactions on the Web
  (TWEB)}, vol.~6, no.~1, p.~2, 2012.

\bibitem{spirin2012survey}
N.~Spirin and J.~Han, ``Survey on web spam detection: principles and
  algorithms,'' \emph{ACM SIGKDD Explorations Newsletter}, vol.~13, no.~2, pp.
  50--64, 2012.

\bibitem{chellapilla2007taxonomy}
K.~Chellapilla and A.~Maykov, ``A taxonomy of javascript redirection spam,'' in
  \emph{Proceedings of the 3rd international workshop on Adversarial
  information retrieval on the web}.\hskip 1em plus 0.5em minus 0.4em\relax
  ACM, 2007, pp. 81--88.

\bibitem{wiki:Marquee}
\BIBentryALTinterwordspacing
Wikipedia, ``Marquee element --- {W}ikipedia{,} the free encyclopedia,'' 2013,
  [Online; accessed 19-January-2013]. [Online]. Available:
  \url{\url{http://en.wikipedia.org/wiki/Marquee\_element}}
\BIBentrySTDinterwordspacing

\bibitem{wiki:Blink}
\BIBentryALTinterwordspacing
``Blink element --- {W}ikipedia{,} the free encyclopedia,'' 2013, [Online;
  accessed 20-January-2013]. [Online]. Available:
  \url{\url{http://en.wikipedia.org/wiki/Blink\_element}}
\BIBentrySTDinterwordspacing

\bibitem{flanagan2006javascript}
D.~Flanagan, \emph{JavaScript: the definitive guide}.\hskip 1em plus 0.5em
  minus 0.4em\relax O'Reilly Media, Incorporated, 2006.

\bibitem{wu2005cloaking}
B.~Wu and B.~Davison, ``Cloaking and redirection: A preliminary study,'' in
  \emph{First International Workshop on Adversarial Information Retrieval on
  the Web (AIRWeb'05)}, 2005.

\bibitem{mori1999optical}
S.~Mori, H.~Nishida, and H.~Yamada, \emph{Optical character recognition}.\hskip
  1em plus 0.5em minus 0.4em\relax John Wiley \& Sons, Inc., 1999.

\bibitem{wkhtmltopdf}
\BIBentryALTinterwordspacing
``wkhtmltopdf,'' 2013, [Online; accessed 20-Febrary-2013]. [Online]. Available:
  \url{\url{http://code.google.com/p/wkhtmltopdf/}}
\BIBentrySTDinterwordspacing

\bibitem{Snapshotter}
\BIBentryALTinterwordspacing
``Snapshotter,'' 2013, [Online; accessed 20-Febrary-2013]. [Online]. Available:
  \url{\url{http://www.mewsoft.com/Products/Snapshotter.html}}
\BIBentrySTDinterwordspacing

\end{thebibliography}

\end{document}